\documentclass[twoside,twocolumn]{IEEEtran}
\usepackage{graphicx}
\usepackage{dcolumn}
\usepackage{array}
\usepackage{amsmath}
\usepackage{fancyheadings}
\usepackage{subfigure}
\usepackage{algorithm}
\usepackage{algorithmic}
\newcolumntype{d}[1]{D{.}{.}{#1}}
\newcolumntype{i}{D{.}{}{0}}

\begin{document}
\pagestyle{empty}
\date{}
\title{Informed Dynamic Scheduling for Belief-Propagation Decoding of LDPC Codes}
\author{Andres I. Vila Casado, Miguel Griot and Richard D. Wesel\\
Department of Electrical Engineering, University of California,
Los Angeles, CA 90095-1594\\Email:
avila@ee.ucla.edu, mgriot@ee.ucla.edu, wesel@ee.ucla.edu
\thanks{This work was supported by the state of California and ST Microelectronics through UC discovery grant
COM 03-10142.}}

\maketitle \thispagestyle{empty}
\begin {abstract}
Low-Density Parity-Check (LDPC) codes are usually decoded by running an iterative belief-propagation, or message-passing, algorithm over the
factor graph of the code. The traditional message-passing schedule consists of updating all the variable nodes in the graph, using the same
pre-update information, followed by updating all the check nodes of the graph, again, using the same pre-update information. Recently several
studies show that sequential scheduling, in which messages are generated using the latest available information, significantly improves the
convergence speed in terms of number of iterations. Sequential scheduling raises the problem of finding the best sequence of message updates.
This paper presents practical scheduling strategies that use the value of the messages in the graph to find the next message to be updated.
Simulation results show that these informed update sequences require significantly fewer iterations than standard sequential schedules.
Furthermore, the paper shows that informed scheduling solves some standard trapping set errors. Therefore, it also outperforms traditional
scheduling for a large numbers of iterations. Complexity and implementability issues are also addressed.
\end {abstract}
\begin {keywords}
Belief propagation, message-passing schedule, error-control codes, low-density parity-check codes.
\end {keywords}
\section {Introduction}

Belief Propagation (BP) provides Maximum-Likelihood (ML) decoding over a cycle-free factor-graph representation of a code as shown in
\cite{mceliece98} and \cite{factorgraph}. In some cases, BP over loopy factor graphs of channel codes has been shown to have near ML
performance. BP performs well on the bi-partite factor graphs composed of variable nodes and check nodes that describe LDPC codes.

However, loopy BP is an iterative algorithm and therefore requires a message-passing schedule. Flooding, or simultaneous scheduling, is the most
popular scheduling strategy.  In every iteration flooding simultaneously updates all the variable nodes (with each update using the same set of
pre-update data) and then, updates all the check nodes (again, with each update using the same pre-update information). Recently, several papers
have addressed the effects of different types of sequential, or non-simultaneous, scheduling strategies in BP LDPC decoding.  The idea was
introduced as a sequence of check-node updates in \cite{yeo01} and \cite{mansourlbp} and as a sequence of variable-node updates in
\cite{kfir03}. It is also presented in \cite{hocevar04} under the name of Layered BP (LBP), in \cite{sharon04} as serial schedule, in
\cite{zhang05} as shuffled BP, in \cite{rado05} as row message passing, column message passing and row-column message passing, among others.

Simulations and theoretical tools in these works show that sequential scheduling converges twice as fast as flooding when used in LDPC decoding.
It has also been shown that sequential updating doesn't increase the decoding complexity per iteration, thus allowing the convergence speed
increase at no cost \cite{rado05}.  In \cite{Guilloud}, where a global framework for the analysis of LDPC decoders is introduced, the complexity
is assumed to be independent from the sequential scheduling chosen. Furthermore, different types of sequential schedules such as sequential
check-node updating, sequential variable-node updating and sequential message updating have very similar performance results \cite{rado05}.
Given their similarities, the different types of sequential updates will be referred in this paper as Standard Sequential Scheduling (SSS).  In
the simulations presented in this paper the SSS strategy used for comparison is LBP, a sequence of check-node updates, as presented in
\cite{mansourlbp} and \cite{hocevar04}.

Sequential updating poses the problem of finding the ordering of message updates that results in the best convergence speed and/or code
performance. The current state of the messages in the graph can be used to dynamically update the schedule, producing what we call an Informed
Dynamic Schedule (IDS) and presented in \cite{andresITA}. To our knowledge, the only well defined informed sequential scheduling is the Residual
Belief Propagation (RBP) algorithm presented by Elidan et al. in \cite{elidan06}. They proposed it for general sequential message passing, not
specifically for BP decoding.

RBP is a greedy algorithm that organizes the message updates according to how different is the message generated in the current iteration from
the message generated in the previous iteration. The intuition is that the bigger this difference, the further from convergence this part of the
graph is. Therefore, propagating this message first will make BP converge at a higher speed.

Simulations show that RBP LDPC decoding has a higher convergence speed than SSS but its error-rate performance for a large enough number of
iterations is worse.  This behavior is commonly found in greedy algorithms, which tend to arrive at a solution faster, but arrive at the correct
solution less often.  We propose a less-greedy IDS in which all the outgoing messages of a check-node are generated simultaneously. We call this
IDS node-wise RBP. It converges both faster and more often than SSS.

Both RBP and node-wise RBP require the knowledge of the message to be updated in order to pick which message to update.  This means that several
messages are computed and not passed. Thus, increasing the complexity of the decoding per iteration.  We propose using the min-BP check-node
update algorithm explained in \cite{chen02} and \cite{jones03} to simplify the ordering metric and significantly decrease the complexity for
both informed scheduling strategies while maintaining the same performance. Also, an analysis of the hardware issues that may arise in a
parallel implementation of these informed sequential scheduling strategies is presented.

This paper is organized as follows. Section \ref{sec:ldpc} reviews LDPC decoding and the flooding and SSS schedules. Section \ref{sec:rbp}
explains how to implement RBP decoding for LDPC codes. This section also introduces and justifies node-wise RBP. Section \ref{sec:hardware}
addresses some complexity and implementability issues. The simulation results of all the message-passing schedules are compared and discussed in
Section \ref{sec:results}. Section \ref{sec:conclusions} delivers the conclusions.

\section {BP decoding for LDPC codes}
\label {sec:ldpc}

In general, BP consists of the exchange of messages between the
nodes of a graph.  Each node generates and propagates messages to
its neighbors based on its current incoming messages.  The vector
of all the messages in the graph is denoted by {\bf m} and $m_k$
denotes the $k$'th message in the vector {\bf m}. The function
that generates $m_k$ from {\bf m} is denoted by ${f_k \left( {\bf
m} \right)}$.

The LDPC code graph is a bi-partite graph composed by $N$ variable nodes $v_j$ for $j \in \{1, \ldots, N\}$ that represent the codeword bits and
$M$ check nodes $c_i$ for $i \in \{1, \ldots, M\}$ that represent the parity-check equations. The exchanged messages correspond to the
Log-Likelihood Ratio (LLR) of the probabilities of the bits. The sign of the LLR indicates the most likely value of the bit and the absolute
value of the LLR gives the reliability of the message.  In this fashion, the channel information LLR of the variable node $v_j$ is $C_{v_j }  =
\log \left( {\frac{{p\left( {y_j \left| {v_j = 0} \right.} \right)}}{{p\left( {y_j \left| {v_j = 1} \right.} \right)}}} \right)$, where $y_j$ is
the received signal. Then, for any $c_i$ and $v_j$ that are connected, the two message generating functions, $m_{v_j  \to c_i }  = {f_{v_j  \to
c_i } \left( {\bf m} \right)}$ and $m_{c_i  \to v_j }  = {f_{c_i \to v_j } \left( {\bf m} \right)}$, are:

\begin{equation} \label{eq:v2c}
m_{v_j  \to c_i } = \sum\limits_{c_a \in {\cal N}\left( {v_j }
\right)\backslash c_i } {m_{c_a \to v_j } }  + C_{v_j },
\end{equation}

\begin{equation} \label{eq:c2v}
m_{c_i  \to v_j }  = 2 \times {\rm atanh}\left( {\prod\limits_{v_b
\in {\cal N}\left( {c_i } \right)\backslash v_j } {\tanh \left(
{\frac{{m_{v_b  \to c_i } }}{2}} \right)} } \right),
\end{equation}
where ${\cal N}\left( {v_j } \right)\backslash c_i$ denotes the neighbors of $v_j$ excluding $c_i$, and ${\cal N}\left( {c_i } \right)\backslash
v_j$ denotes the neighbors of $c_i$ excluding $v_j$.

BP decoding consists of the iterative update of the messages until
a stopping rule is satisfied.  In flooding scheduling, an
iteration consists on the simultaneous update of all the messages
$m_{v \to c }$ followed by the simultaneous update of all the
messages $m_{c \to v }$. In SSS, an iteration consists on the
sequential update of all the messages $m_{v \to c }$ as well as
all the messages $m_{c \to v }$ in a specific pre-defined order.
The algorithm stops if the decoded bits satisfy all the
parity-check equations or a maximum number of iterations is
reached.

\section {Residual Belief Propagation (RBP)}
\label {sec:rbp}

RBP, as introduced in \cite{elidan06}, is an informed scheduling strategy that updates first the message that maximizes an ordering metric
called the residual. A residual is the norm (defined over the message space) of the difference between the values of a message before and after
an update. For a message $m_k$, the residual is defined as:

\begin{equation} \label{eq:residual}
r\left( m_k \right) = \left\| {f_k \left( {\bf m} \right) - m_k }
\right\|.
\end{equation}

The intuitive justification of this approach is that as loopy BP converges, the differences between the messages before and after an update go
to zero. Therefore if a message has a large residual, it means that it's located in a part of the graph that hasn't converged yet.  Therefore,
propagating that message first should speed up the process.  Elidan \emph{et al.} create a priority queue, ordered by the value of the residual,
so at each step the first message in the queue is updated and then the queue is reordered using the new information.

\vspace{-.1in}
\subsection{RBP decoding for LDPC codes}

In LLR BP decoding, all the message spaces are one-dimensional
(the real line). Therefore, the residuals are the absolute values
of the difference of the LLRs.

Let us analyze the behavior of RBP decoding for LDPC codes in order to simplify the decoding algorithm. Initially, all the messages $m_{v \to
c}$ are set to the value of their correspondent channel message $C_v$. No operations are needed in this initialization. This implies that the
residuals of all the variable-to-check messages $r(m_{v \to c})$ are equal to 0. Then, without loss of generality, we assume that the message
$m_{c_i \to v_j }$ has the residual $r^*$, which is the biggest of the graph. After $m_{c_i \to v_j }$ is propagated, only residuals $r(m_{v_j
\to c_a})$ change, with $c_a \in {\cal N}\left( {v_j } \right)\backslash c_i$.

The new residuals $r(m_{v_j \to c_a})$ are equal to $r^*$, because
$r^*$ was the change in the message $m_{c_i \to v_j }$ and Eq.
\ref{eq:v2c} shows that the message update operations of a
variable node are only sums. Therefore, the messages $m_{v_j \to
c_a}$ have now the biggest residuals in the graph.

Assuming that propagating the messages $m_{v_j \to c_a}$ won't
generate any new residuals bigger than $r^*$, RBP can be
simplified. Every time a message $m_{c \to v}$ is propagated, the
outgoing messages from the variable node $v$ will be updated and
propagated. This facilitates the scheduling since we need only to
maintain a queue $Q$ of messages $m_{c \to v}$, ordered by the
value of their residuals, in order to find out the next message to
be propagated. RBP LDPC decoding is formally described in
Algorithm \ref{alg:rbp}, the stopping rule will be discussed in
Section \ref{sec:conclusions}.

\begin{algorithm}[h]
\caption{RBP decoding for LDPC codes}\label{alg:rbp}
\begin{algorithmic} [1]
\STATE  Initialize all $m_{c \to v}=0$ \STATE Initialize all
$m_{v_j \to c_i}=C_j$ \STATE Compute all $r(m_{c \to v})$ and
generate $Q$ \STATE Let $m_{c_i  \to v_j }$ be the first message
in $Q$ \STATE Generate and propagate $m_{c_i \to v_j }$ \STATE Set
$r(m_{c_i  \to v_j })=0$ and re-order $Q$ \FOR{every $c_a \in
{\cal N}\left( {v_j } \right)\backslash c_i$ }
    \STATE Generate and propagate $m_{v_j  \to c_a}$
    \FOR{every $v_b \in {\cal N}\left( {c_a } \right)\backslash v_j$ }
        \STATE Compute $r(m_{c_a  \to v_b})$ and re-order $Q$
    \ENDFOR
\ENDFOR \IF{Stopping rule is not satisfied} \STATE Position=4;
\ENDIF
\end{algorithmic}
\end{algorithm}

There is an intuitive way to see how RBP decoding works for LDPC codes.  Let us assume that at a certain moment in the decoding, there is a
check node $c_i$ with residuals $r(m_{c_i \to v_b})=0$ for every $v_b \in {\cal N}\left( {c_i } \right)$. Now let us assume that there is a
change in the value of the message $m_{v_j \to c_i}$. The biggest change in a check-to-variable message out of $c_i$ (therefore the largest
residual) will happen in the edge that corresponds to the incoming variable-to-check message with the lowest reliability (excluding the message
$m_{v_j \to c_i}$). Let us denote by $v_k$ the variable node that is the destination of the message that has the largest residual $r(m_{c_i \to
v_k})$.  Then, the message $m_{v_k \to c_i}$ has the smallest reliability out of all messages $m_{v_b \to c_i}$, with $v_b \in {\cal N}\left(
{c_i } \right)\backslash v_j$.

This implies that, for this particular scenario, once there's a change in a variable-to-check message, RBP will propagate first the message to
the variable node with the lowest reliability. This makes sense intuitively. In some sense the lowest reliability variable node needs to receive
new information more than the higher reliability ones.

The negative effects of the greediness of RBP are apparent in the case of unsatisfied check nodes. RBP will schedule to propagate first the
message that will ``correct" the variable node with the lowest reliability. This is the most likely variable node to be in error. However, if
that variable node was already correct, changing its sign will likely generate new errors, making the BP convergence more difficult. This
analysis helps us see why RBP corrects the most likely errors faster but has trouble correcting ``difficult" errors as will be seen in Section
\ref{sec:results}. We define ``difficult" errors as the errors that need a large number of message updates to be corrected.

\vspace{-.1in}
\subsection{Node-wise RBP decoding for LDPC codes}

In order to obtain a better performance for all types of errors, perhaps a less greedy scheduling strategy must be used. As noted earlier, some
of the greediness of RBP came from the fact that it tends to propagate first the message to the less reliable variable nodes. We propose to
simultaneously update and propagate all the check-to-variable messages that correspond to the same check node $c_i$, instead of only propagating
the message with the largest residual $r(m_{c_i  \to v_j })$. It can be seen, using the analysis presented earlier, that this algorithm is less
likely to generate new errors. We call this less greedy strategy node-wise RBP and it's performance can be seen in Section \ref{sec:results}.
Node-wise RBP is similar to LBP; it is a sequence of check-node updates. However, unlike LBP, which follows a pre-determined order, the check
node to be updated next is chosen dynamically according to the residuals of the check-to-variable messages.  Node-wise RBP is formally described
in Algorithm \ref{alg:nwrbp}.

\begin{algorithm}[h]
\caption{Node-wise RBP decoding for LDPC codes}\label{alg:nwrbp}
\begin{algorithmic} [1]
\STATE  Initialize all $m_{c \to v}=0$\STATE Initialize all
$m_{v_j \to c_i}=C_j$  \STATE Compute all $r(m_{c \to v})$ and
generate $Q$ \STATE Let $m_{c_i  \to v_j }$ be the first message
in $Q$ \FOR{every $v_k \in {\cal N}\left( {c_i } \right)$ } \STATE
Generate and propagate $m_{c_i \to v_k}$ \STATE Set $r(m_{c_i  \to
v_k})=0$ and re-order $Q$ \FOR{every $c_a \in {\cal N}\left( {v_k
} \right)\backslash c_i$ }
    \STATE Generate and propagate $m_{v_k  \to c_a}$
    \FOR{every $v_b \in {\cal N}\left( {c_a } \right)\backslash v_k$ }
        \STATE Compute $r(m_{c_a  \to v_b})$ and re-order $Q$
    \ENDFOR
\ENDFOR \ENDFOR \IF{Stopping rule is not satisfied} \STATE
Position=4; \ENDIF
\end{algorithmic}
\end{algorithm}

Node-wise RBP converges both faster than SSS (in terms of number of messages updated) and better than SSS (in terms of FER of the code for a
large number of iterations).  We can explain intuitively and demonstrate experimentally that the lower error rates are achieved because informed
scheduling allows the LDPC decoder to overcome many "trapping sets". Trapping sets, or near-codewords, as defined in \cite{richardsonef}, are
small variable-node sets such that the induced sub-graph has a small number of odd degree neighbors. In \cite{richardsonef}, Richardson also
mentions that the most troublesome trapping set errors are those where the odd degree neighbors have degree 1 (in the induced sub-graph), and
the even-degree neighbors have degree 2 (in the induced sub-graph).

It is likely that node-wise RBP solves the variable nodes in error by sequentially updating the degree-1 check-nodes connected to them. When a
variable node in a trapping set is corrected, the induced sub-graph of the variable-nodes-in-error will change as follows. At least one check
node that was degree-2, becomes degree-1 after the variable node correction. This check node is likely to be picked as the next check-node to be
updated by node-wise RBP because its messages will have large residuals. This update will probably correct another variable node in the trapping
set.

We corroborated this analysis by studying the behavior of the decoders for the noise realizations that the SSS decoder could not solve for 200
iterations and that node-wise RBP solved in a very small number of iterations (less than 10).  We found that a large majority of the SSS errors
in these cases are caused by trapping sets that node-wise RBP solved in the manner mentioned before.

\section {Complexity and Implementation}
\label {sec:hardware}

Given the surge in popularity of LDPC codes for practical implementations, it is interesting to address some issues about the complexity of RBP
and node-wise RBP when compared to SSS and flooding. \vspace{-.1in}

\subsection{Complexity per iteration}

For an IDS we consider one iteration to have occurred after the number of updates equals the number of updates completed in an SSS or flooding
iteration. For RBP, an iteration will be counted after the number of check-to-variable message updates equals the number of edges in the LDPC
graph. For node-wise RBP an iteration will be counted after a number of check-node updates equals the number of check-nodes of the code.

In \cite{rado05}, the authors prove that if the appropriate update
equations are chosen, the total number of operations per iteration
of all the sequential schedules is equal to the number of
operations per iterations of the flooding schedule, making their
complexity per iteration equal. Given that both RBP and node-wise
RBP are forms of sequential updates, a sequence of message updates
in the first case and a sequence of check-node updates in the
second, then they also use, on average, the same number of
message-generating operations per iteration (using our definition
of iteration for informed schedules).

In order to maintain the same complexity, we use the typical
stopping rules in the decoding. Stop if at the end of an iteration
the decoded bits satisfy all the parity-check equations, or a
maximum number of iterations is reached.

On top of the message-generation complexity, informed schedules incur two extra processes: residual computation and ordering of the residuals.
As defined in Section \ref{sec:rbp}, the residual computation requires the value of the message that would be propagated. This requires
additional complexity since there will be numerous message computations that will only be used to calculate residuals.  We propose to use the
min-BP check-node update approximation explained in \cite{chen02} to compute the approximate-residual as follows,

\begin{equation} \label{eq:aresidual}
r^{(a)}\left( m_k \right) = \left\| {f_k^{(a)} \left( {\bf m}
\right) - m^{(a)}_k } \right\|,
\end{equation}
where the superscript $(a)$ stands for approximate and indicates the min-BP approximation.  The min-BP check-node update consists of finding the
two variable-to-check messages with the smallest reliability. Then, the smallest reliability is assigned as the check-to-variable message
reliability for all the edges except the one where the smallest reliability came from.  The second smallest reliability is assigned to that
remaining edge.  The proper sign is computed for all the check-to-variable messages.  Thus, replacing all the residual functions for
approximate-residual functions in Algorithms \ref{alg:rbp} and \ref{alg:nwrbp}, defines Approximate RBP (ARBP) decoding and node-wise ARBP
decoding. These significantly simpler algorithms perform as well as the ones presented in Section \ref{sec:rbp}, as will be seen in Section
\ref{sec:results}. Note that we only propose to use min-BP for the residual computation. For the actual propagation of messages we use the full
update equations (\ref{eq:v2c}) and (\ref{eq:c2v}).

Even for node-wise ARBP, the practical trade-off between the increase in the per-iteration complexity and the decrease in the number of
iterations (and improved FER) is difficult to address in general as it depends on specific implementation choices. Our current research is
addressing this trade-off in detail.

\vspace{-.1in}

\subsection{Parallel Decoding}

The possibility of having several processors computing messages at the same time during the LDPC decoding has become an intense area of research
and an important reason why LDPC codes are so successful. Furthermore, codes with a specific structure have been shown to allow SSS decoding
while maintaining the same parallelism degree obtained for flooding decoding \cite{mansourlbp}.  In principle, the idea of having an ordered
sequence of updates, that uses the most recent information as much as possible, isn't compatible with the idea of simultaneously computing
messages.  However, since the ordering of the queue $Q$ based on the new results occurs after the update, it is possible that the several
parallel processors can work on different parts of the graph while still using the most recent information.

We define the parallel node-wise ARBP scheduling strategy as the node-wise ARBP strategy where instead of updating only the check-node with the
largest approximate residual, $p$ check-nodes are updated at the same time. The $p$ nodes that have the largest approximate residuals are
updated simultaneously. These $p$ check nodes are not designed to work in parallel, unlike the $p$ check-nodes of a $p \times p$ sub-matrix as
defined in \cite{mansourlbp}.

However, parallel processing may be implemented extending the
hardware solutions presented in \cite{rovini06}.  For instance, if
one or more check-nodes have in common one or more variable nodes,
they will all use the same previous information and compute the
incremental variations that are afterwards combined in the
variable-node update.  There are hardware issues, such as memory
clashes, that still need to be carefully addressed when
implementing parallel node-wise ARBP.

Parallel node-wise ARBP has a very small performance degradation when compared with node-wise ARBP, as will be seen in Section
\ref{sec:results}. We  defined and simulated the parallel version of node-wise ARBP since it's the simplest IDS strategy and therefore, the most
likely to be implemented.


\section {Simulation Results}
\label {sec:results}

This section presents the AWGN performance of the different scheduling strategies presented above.  All the simulations are floating point and
use the same rate-1/2 LDPC code.  The blocklength of the code is 1944 and it has the same sub-matrix structure as the one presented in
\cite{mansourlbp} with sub-matrix size equal to 54x54.

The SSS results shown correspond to the sequential check-node
update introduced in \cite{mansourlbp}.  This scheduling is known
as Layered Belief Propagation (LBP) and guarantees a parallelism
degree equal to the sub-matrix size of the LDPC code (54 in our
case). As shown in \cite{rado05}, different SSS strategies produce
almost identical results so its selection doesn't significantly
affect the performance of the decoder.

\begin{figure}
\begin{center}
\scalebox{1}{\includegraphics{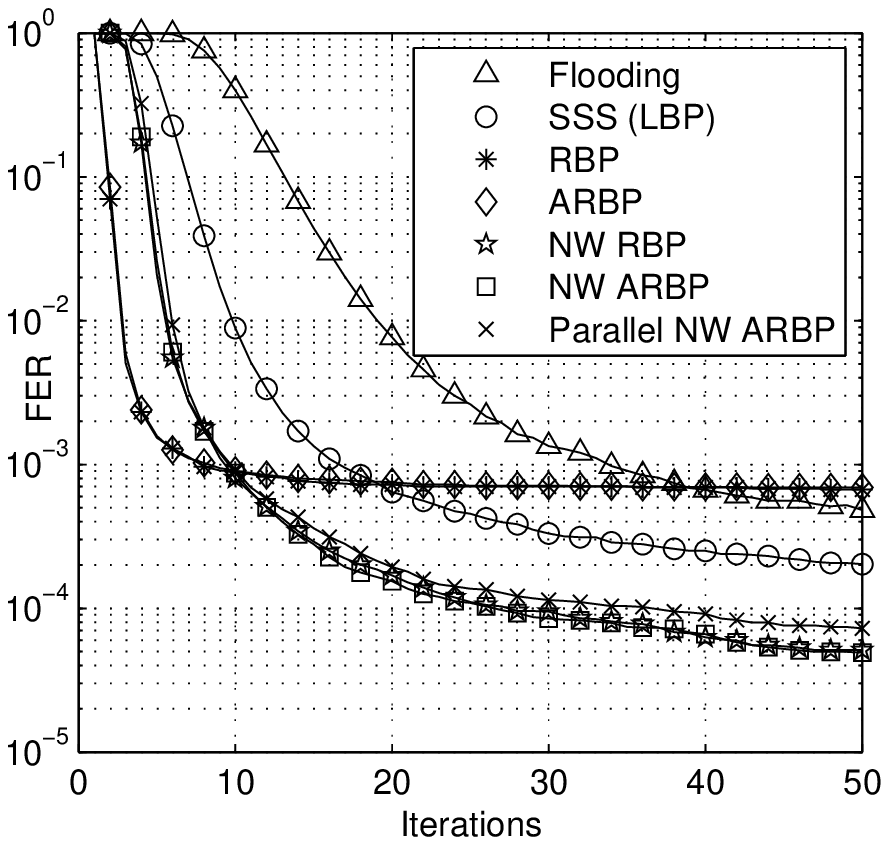}} \caption{FER Performance of flooding, SSS (LBP), RBP and node-wise RBP vs. number of iterations
for a fixed $E_b/N_o=1.75$ dB} \label{fig:all}\end{center} \vspace{-.2in}
\end{figure}

Fig. \ref{fig:all} shows the performance of the scheduling strategies discussed above, flooding, SSS (LBP), RBP, ARBP, node-wise RBP, node-wise
ARBP, and parallel node-wise ARBP, as the number of iterations increases.  The figure shows that RBP has a significantly better performance than
SSS (LBP) for a small number of iterations, but a sub-par performance for a larger number of iterations.  Specifically, the performance of RBP
at 4 iterations is equal to the performance of SSS (LBP) at 13 iterations, but the curves cross over at 19 iterations.  This suggests that RBP
has trouble with ``difficult" errors as discussed earlier.

Node-wise RBP, while not as good as RBP for a small number of iterations, shows consistently better performance than SSS (LBP) across all
iterations.  Specifically, the performance of node-wise RBP at 18 iterations is equal to the performance of SSS (LBP) at 50 iterations. The
results for flooding are shown for comparison purposes, and replicate the theoretical and empirical results of \cite{mansourlbp}-\cite{rado05}
that claim that flooding needs twice the number of iterations as SSS.

Fig. \ref{fig:all} also shows the performance of the approximate residual schedules and compares them with the schedules that use the exact
residuals. It can be seen that both ARBP and node-wise ARBP perform almost indistinguishably from RBP and node-wise RBP respectively.  We
reiterate that the approximate residual diminishes the complexity of residual computation significantly, thus making ARBP, and node-wise ARBP
more attractive than their exact counterparts.


Furthermore, Fig. \ref{fig:all} also shows the performance of parallel node-wise ARBP.  The relatively small loss in performance when compared
to node-wise ARBP is the price for the throughput increase resulting from the parallelism. The number $p$ of check-nodes processed in parallel
was set to 54, which is equal to the parallelism guaranteed by SSS (LBP) decoding this structured LDPC code with sub-matrix size equal to 54x54
\cite{mansourlbp}.

\begin{figure}
\begin{center}
\scalebox{1}{\includegraphics{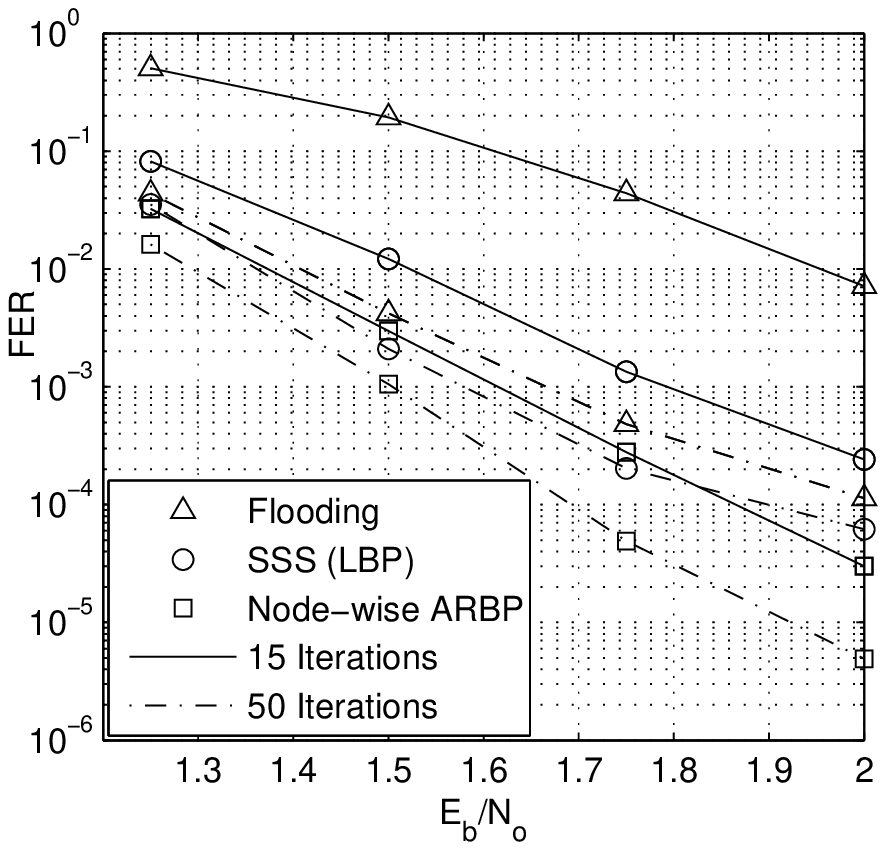}} \caption{FER Performance of flooding, SSS (LBP) and node-wise ARBP for 15 and 50 iterations vs.
$E_b/N_o$} \label{fig:nwarbp}\end{center} \vspace{-.2in}
\end{figure}

The FER of node-wise ARBP vs. SNR and for 15 and 50 iterations (maximum) is presented in Fig. \ref{fig:nwarbp}. The FER of flooding and SSS
(LBP) are also presented as references.  It can be seen that the SNR gap between node-wise ARBP and SSS (LBP) is more pronounced for a small
number of iterations and/or a large SNR.

\begin{figure}
\begin{center}
\scalebox{1}{\includegraphics{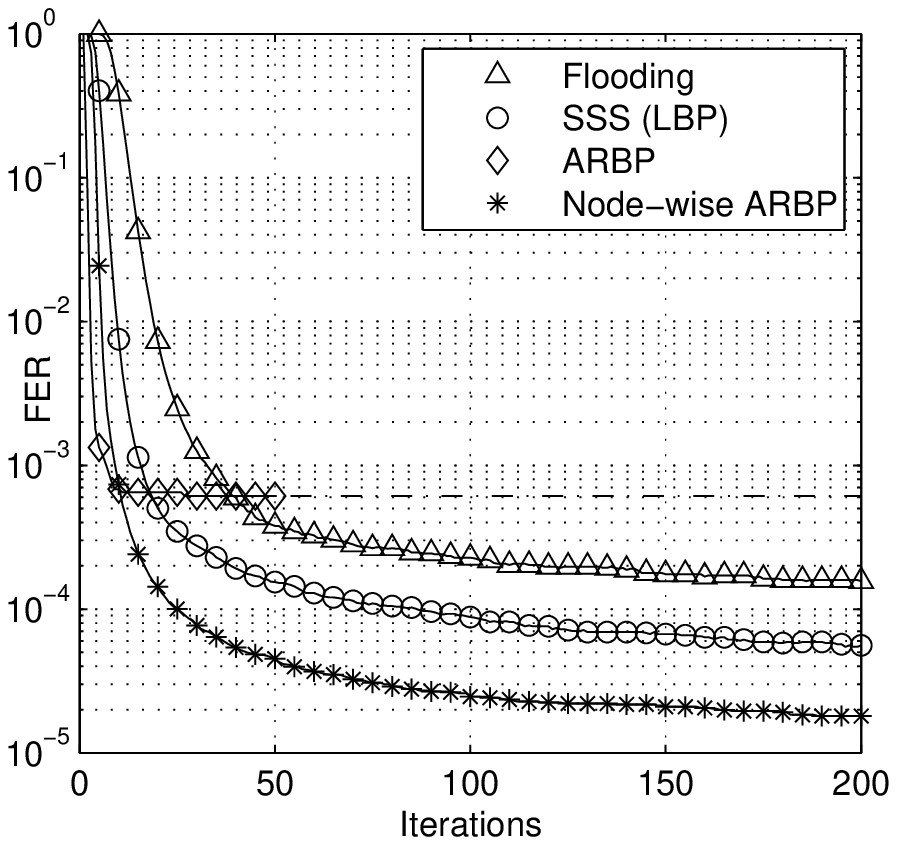}} \caption{FER vs. number of iterations of the 802.11n blocklength-1944 rate-1/2 code using
flooding, SSS (LBP), ARBP and node-wise ARBP for a fixed $E_b/N_o=1.75$ dB} \label{fig:1211n}\end{center} \vspace{-.2in}
\end{figure}

\begin{figure}
\begin{center}
\scalebox{1}{\includegraphics{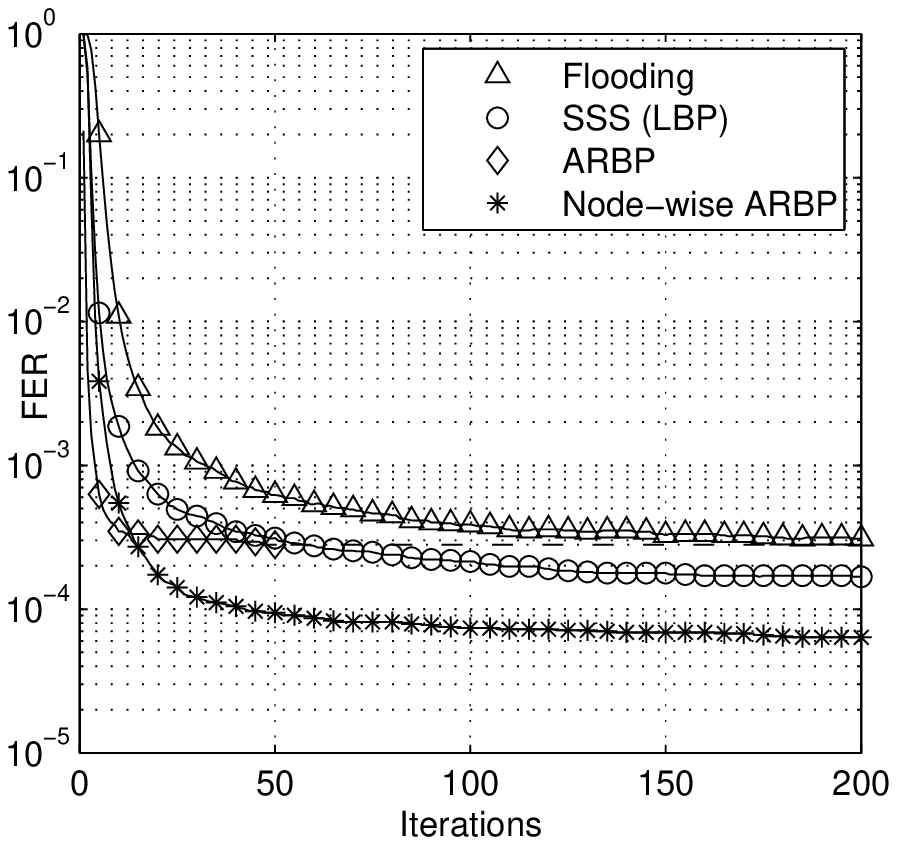}} \caption{FER vs. number of iterations of the 802.11n blocklength-1944 rate-5/6 802.11n code
using flooding, SSS (LBP), ARBP and node-wise ARBP for a fixed $SNR=6.0$ dB} \label{fig:5611n}\end{center} \vspace{-.2in}
\end{figure}

Fig. \ref{fig:1211n} and Fig. \ref{fig:5611n} show the performance of different scheduling strategies for the blocklength 1944 rate-1/2 and rate
5/6 LDPC codes selected for the IEEE 802.11n standard \cite{IEEE80211n}.  These simulations were run for a high number of iterations (200) and
show that node-wise ARBP achieves a better FER performance that SSS (LBP).  Fig. \ref{fig:5611n} also shows that even for high rate codes,
node-wise ARBP converges both faster and better than SSS (LBP).

\section {Conclusions}
\label {sec:conclusions}

This paper shows that, while maintaining the same message-generation functions, IDS can improve the performance of BP LDPC decoding.

RBP and its simplification ARBP are appropriate for applications that have a high target error-rate, given that RBP achieves these error-rates
using significantly fewer iterations than SSS. They are also appropriate for high-speed applications that only allow a small number of
iterations.  However, for applications that require lower error rates and allow larger delays RBP and ARBP aren't appropriate.

For such applications node-wise RBP and its simplification node-wise ARBP perform better than SSS for any target error-rate and any number of
iterations. These node-wise strategies achieve a lower error-rates by overcoming trapping set errors that SSS cannot solve. Furthermore a
parallel implementation of node-wise ARBP was shown to perform nearly as well as the original node-wise ARBP, making this informed scheduling
more attractive for practical implementations.

The improvement in performance of these informed scheduling strategies were also shown for a high-rate code (rate 5/6).  However, they come with
the cost of an increase in complexity per iteration due to the residual computation and its ordering. The  trade-off provided by node-wise ARBP
between increasing the per-iteration complexity and reducing the number of iterations (while also reducing the FER for a large number of
iterations) requires further investigation in the context of specific implementations.

The ideas presented in this work may be extended to other communication solutions that use iterative BP, such as turbo codes,
turbo-equalization, iterative demodulation and decoding of high-order constellations.  The extensions of the IDS strategies may also prove
beneficial for loopy BP solutions to problems outside the communications field.

\bibliographystyle{unsrt}
{\bibliography{sched}}

\end{document}